# Categorisation of Spreadsheet Use within Organisations, Incorporating Risk: A Progress Report

**Mukul Madahar,** Pat Cleary, David Ball
Cardiff School of Management
University of Wales Institute Cardiff
Colchester Avenue
Cardiff, CF23 9XR
mmadahar@uwic.ac.uk, pmcleary@uwic.ac.uk, dball@uwic.ac.uk

**ABSTRACT:**

There has been a significant amount of research into spreadsheets over the last two decades. Errors in spreadsheets are well documented. Once used mainly for simple functions such as logging, tracking and totalling information, spreadsheets with enhanced formulas are being used for complex calculative models. There are many software packages and tools which assist in detecting errors within spreadsheets. There has been very little evidence of investigation into the spreadsheet risks associated with the main stream operations within an organisation. This study is a part of the investigation into the means of mitigating risks associated with spreadsheet use within organisations. In this paper the authors present and analyse three proposed models for categorisation of spreadsheet use and the level of risks involved. The models are analysed in the light of current knowledge and the general risks associated with organisations.

1.     INTRODUCTION:

Many organisations rely on spreadsheets as a key tool in their financial reporting and operational processes (Price Waterhouse Coopers (PWC), 2004). Spreadsheets are becoming a significant part of the organisational decision-making framework and there is ample evidence that spreadsheets are erroneous. Panko and Nicholas (2005) highlighted the main studies undertaken to analyse the extent of spreadsheet error, which included studies by Hicks (1995), Coopers and Lybrand (1997), KPMG (1998), Butler (2000), and others. There are many software packages and tools such as SpACE, ComplyXL, ClusterSeven and Actuate which tend to assist in locating and rectifying errors within complex spreadsheet models. Majority of the investigations concerning spreadsheet problems have been from a technical perspective. This paper is a part of the investigation within University of Wales Institute Cardiff (UWIC) that is examining the spreadsheet risks from a management/organisational point of view rather than from a technical stance. The purpose is to relate the risks associated with spreadsheet use to organisational risk management. The aim is to develop a framework for categorising spreadsheet use, incorporating risk as one of the criteria. Once this framework is developed, it could prove very valuable for managing risks associated with spreadsheet use.

Many researchers have categorised errors in spreadsheets (PWC, 2004; Panko, 1996; Galletta, 1993; Rajalingham, Chadwick and Knight, 2000; Rajalingham, 2005). PWC (2004) categorised spreadsheets by use and complexity. This research attempts to incorporate the risk factor into the categorisation of use of spreadsheets. The paper proposes a number of models of categorisation which are based on a pilot study conducted within the accommodation office of the Facilities

Department at UWIC. Subsequently a comparative analysis of the various models proposed is presented.

This paper is organised into three further sections. Section 2 sets out the context of the pilot study. Section 3 highlights the three models that were proposed as a result of the pilot study. Section 4 then presents a comparison of the models proposed. Then finally the conclusions are presented.

## 2. THE STUDY:

### 2.1 Organisation:

For conducting the pilot study the Accommodation Office, was selected within the Facilities Department of UWIC. The Office is responsible for arranging accommodation for students and therefore includes all the halls of residence as well as assisting in finding private accommodation. During the summer, the same department works with the Conference Services Department to organise accommodation for conferences. The spreadsheet use within the department is diverse ranging from basic record keeping (trivial) to complex budgeting and forecasting (strategic).

### 2.2 Methodology:

The research methodology behind the pilot study reflects an interpretive approach due to the nature of this investigation. As Saunders *et al.* (2006) highlights, business situations are complex and unique. This study was conducted in order to understand how this department use spreadsheets and perception of their importance and the associated risks.

As this study was a pilot, it was decided that Un-structured interviews were appropriate, in order to extract maximum information without any bias or influence on the interviewee. As highlighted by Bell (1993), 'unstructured interviews centred round the topic area may produce a wealth of valuable data.' The rationale behind the pilot investigation was to identify and analyse which aspects to explore further and areas that need to be discarded.

The data gathered was analysed to identify the spectrum of use of spreadsheets within the department. The models proposed are based on conducting in-depth one-to-one interviews of all the key personnel involved with development and use of spreadsheets within the department. All the interviews were recorded and then transcribed and analysed.

The authors initially presented the models INFORMS, 2006 and the feedback from that has been incorporated into the discussion.

## 3. FINDINGS:

The spreadsheets were an integral part of the case study department. On one extreme they were used for designing basic paper forms or data stores, whereas on the other extreme they were used for budgeting and forecasting, which formed the more strategic use of spreadsheets.

### 3.1 Model 1:

*Dimensions for Model 1 (Ref. Figure 1):*

- Use: The categories for this dimension of the model were based on the type of use of the spreadsheet. The use of spreadsheets within the department was significant. They were mainly used as a data source. Some of the employees used complex models too, for budgeting and forecasting. Therefore the spreadsheets could be classified as being 'Trivial' data sources or 'Strategic' spreadsheets. Further analysis of the interviews revealed that within the Strategic spreadsheets, there were calculative (The spreadsheets used for budgeting and forecasting) and non-calculative spreadsheets (For example, The UWIC Rider Bible which recorded the details of the student passes for UWIC Bus service). Therefore the first model had three categories within use.

1. Trivial Informative
2. Strategic (Non Calculative/ Informative)
3. Strategic Calculative.

- Importance: This dimension was based purely on reliance on the information contained within the Spreadsheet. Therefore the three categories produced:

1. Not important/ Little Important
2. Important
3. Critical

- Risk: For this dimension the conventional categorisation was used. Any definition of risk is likely to carry an element of subjectivity, depending upon the nature of the risk and to what it is applied. (Riabokon, 2004) As such there is no all encompassing definition of risk. Chicken & Posner (1998) acknowledge this, and instead provide their interpretation of what a risk constituents:

Risk = Hazard x Exposure

They define hazard as "... the way in which a thing or situation can cause harm," (ibid) and exposure as "…The extent to which the likely recipient of the harm can be influenced by the hazard" (ibid). Therefore risk in this case can be assessed on probability of information being incorrect or erroneous and the extent of problems that can be caused by it being erroneous. The problems or the loss can be financial, operative or administrative. Risk, therefore, is measured in terms of impact and likelihood. Therefore the three categories proposed for this model were:

1. Low
2. Medium
3. High

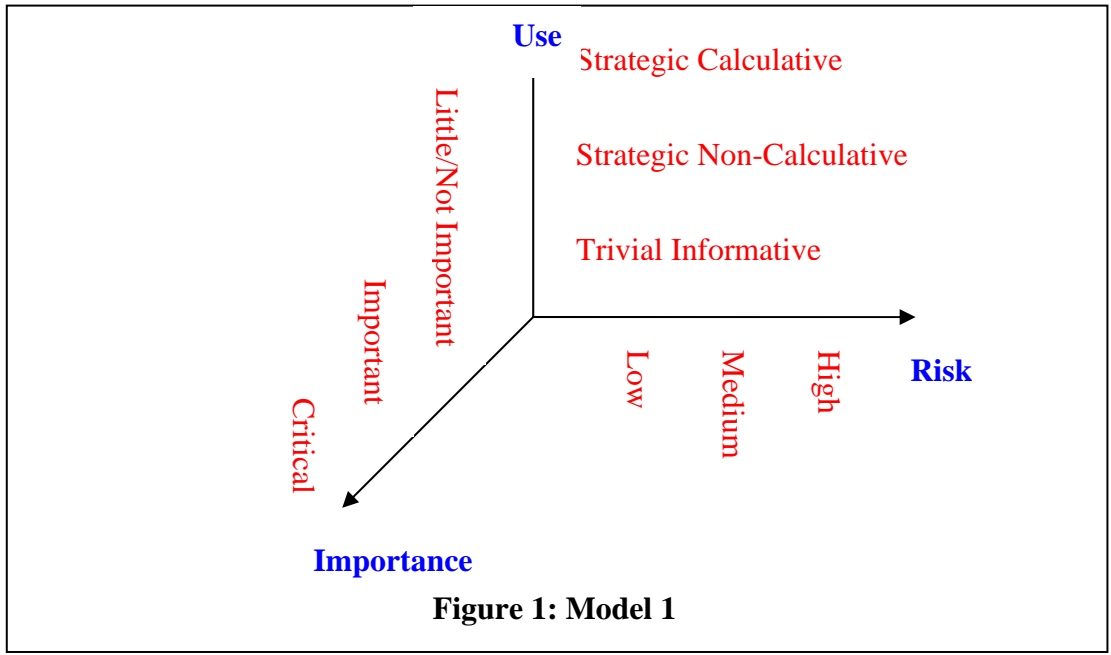

**Figure 1: Model 1**

*Discussion for Model 1 (Refer: Figure 1):*

This model highlights 27 different categories (3 X 3 X 3 Matrix) of spreadsheet usage. The research group at UWIC discussed the above model and it was highlighted that there is some overlap within the categories of 'Use'. Thus it was proposed that it might be worth reducing the use categories to Calculative and Non-Calculative and the spreadsheets that are used for strategic purposes can be categorised on the level of 'Importance'. Thus a new framework for categorisation was proposed (Refer: Figure 2), with just two categories of 'Use' i.e. Calculative and Non-calculative.

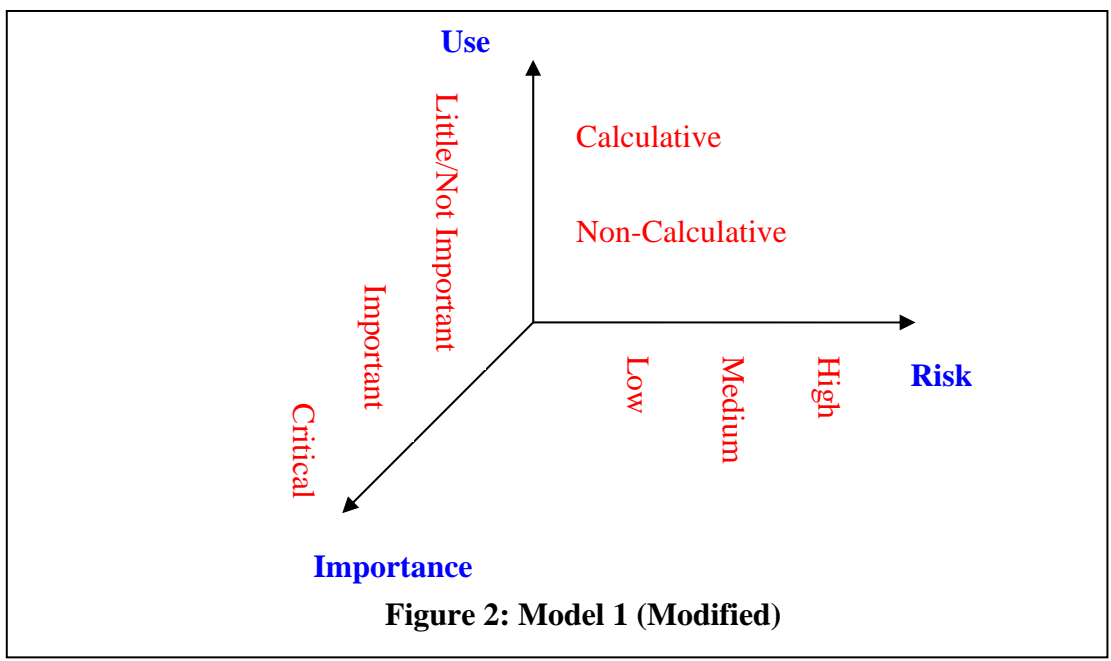

**Figure 2: Model 1 (Modified)**

*3.2  Model 2*

This model was considered to be acceptable specific to the case study department for the pilot but the classification was considered more complex. As can be seen there seems to be some overlap between 'Risk' and 'Importance'. For example, something that represents a risk can be assumed to be important to the organisation. Considering the complexity of the above model, another classification has been proposed (Refer: Figure 3). This model emphasise Importance *vs.* Urgency.

*Dimensions for Model 2 (Refer: Figure 3):*

- The Importance means that dealing with that spreadsheet results in a high pay-off.
- Urgent means tight deadlines are associated with it. (The urgency dimension was inspired by Pawel J. Kalczynski (2006) paper presented at UKAIS Conference 2006.) One of the reasons why spreadsheets are commonly used is that they are quick and easy to use tools. Most of the organisations can easily differentiate what is urgent and what is not.

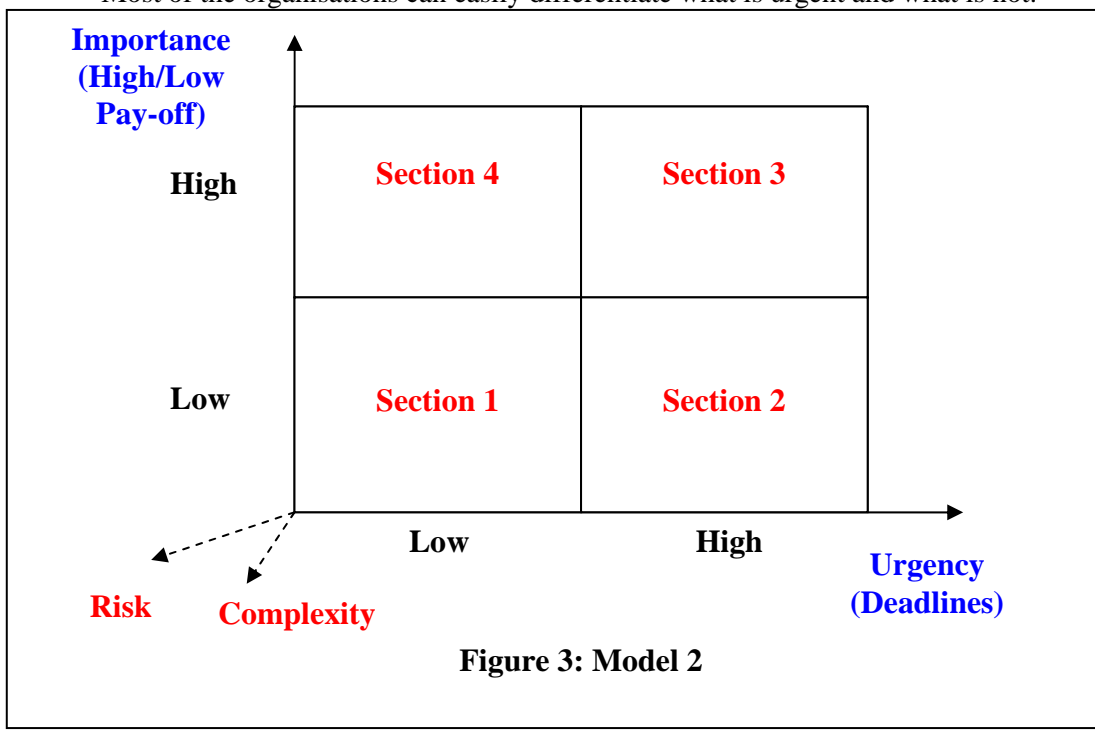

**Figure 3: Model 2**

*Discussion:*

This section now describes the individual sections of the model along with the proposed approach that can be followed for these sections.

- **Section 1:** Low Importance and Low Urgency: Do nothing
- **Section 2:** Low Importance and High Urgency: This is the situation in which spreadsheets, being the quick and easy-to-use tool, are commonly used.
- **Section 3:** High Importance and High Urgency: This is the category of spreadsheet is critical and the organisation requires to put management controls in place.

- **Section 4:** High Importance and Low Urgency: In this category, as the urgency is low, it might be better to use other methods, like database etc.

This model is very simple to implement and apply on individual basis, but there are further two dimensions, which need to be addressed, i.e. 'Complexity' and 'Risk', thereby making it complex and hard to implement in big organisations.

Risk in this model is weighted by assessing management structures in place for three variables: (Refer: Figure 4)

1. Spreadsheet Error
2. Control
3. Compliance

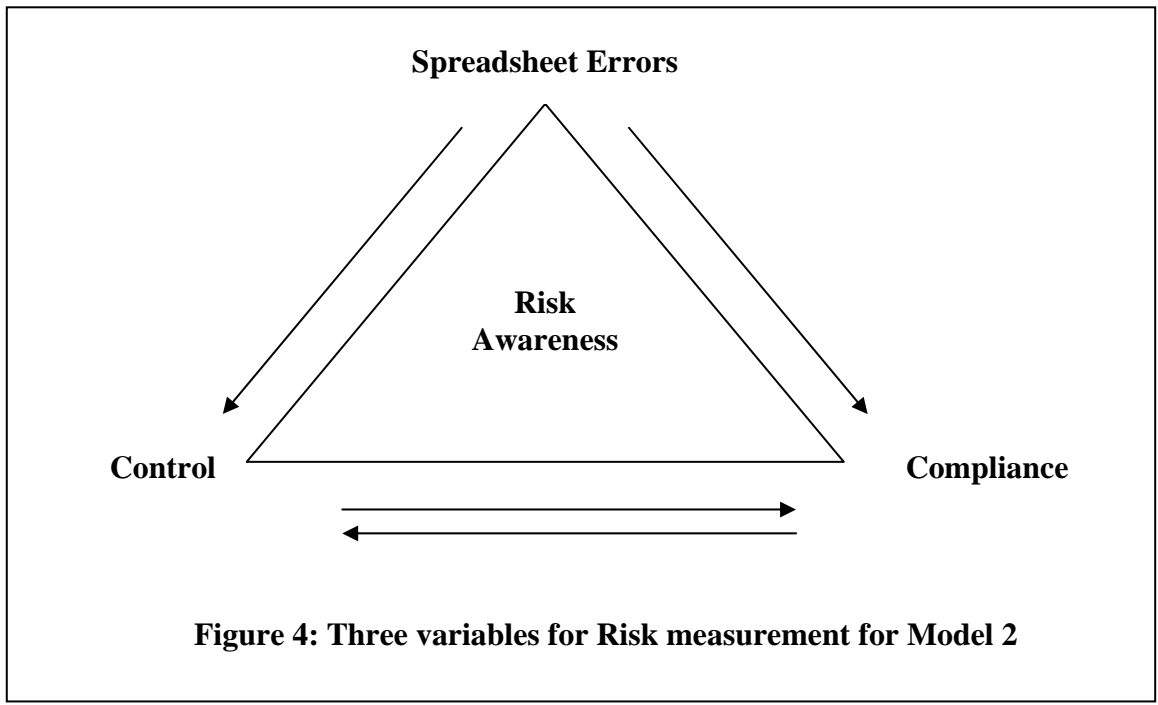

**Figure 4: Three variables for Risk measurement for Model 2**

*3.3    Model 3*

The overlap between the 'Use' and 'Importance' in Model 1, the fuzziness between the 'Risk' and 'Importance' in Model 1 (modified), and problems with implementing Model 2 in big and complex organisations, lead the researcher to propose another model **(Model 3: Refer Figure 5).** This model works on more conventional approaches towards Risk.

*Dimensions for Model 3 (Refer: Figure 5):*

- The X-Axis is to classify the 'Magnitude' of risk i.e. in other words the classification for this dimension can be based on severity of the consequences of errors within spreadsheets, which can be financial or business risk (which can also include reputation and compliance).
- The Y axis highlights the 'Dependency', which can be operational, tactical or strategic.
- The 'Urgency' dimension from Model 2 can also be incorporated into this model.

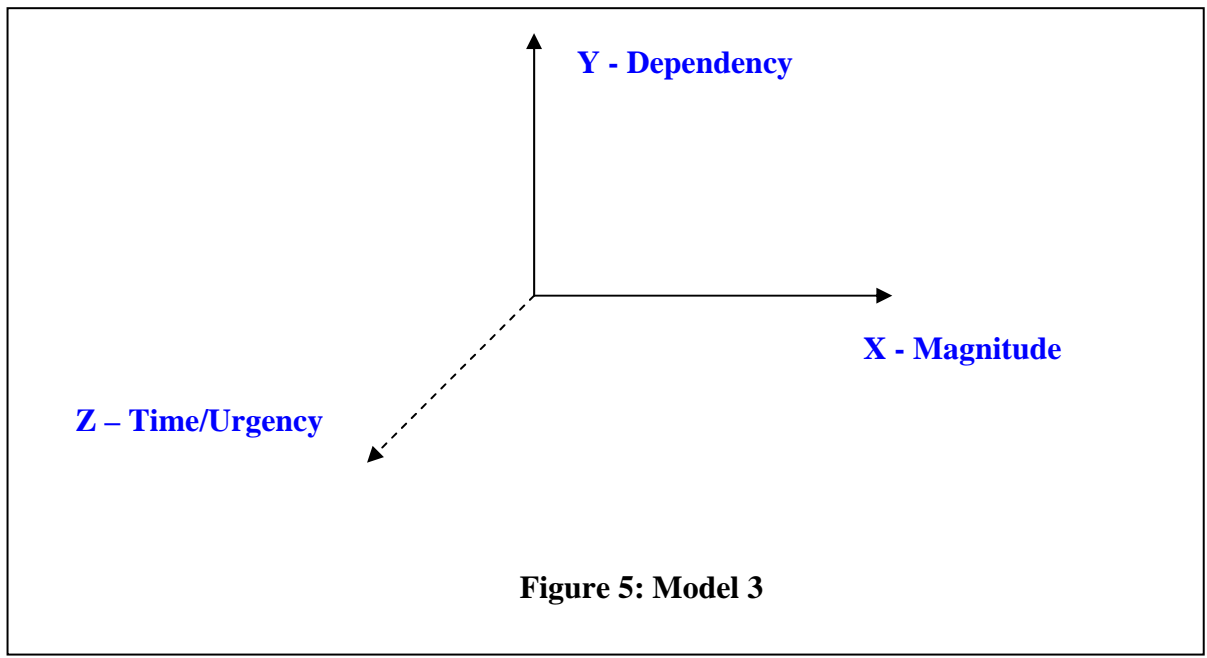

**Figure 5: Model 3**

*Discussion:*
When defining the 'Dependency' dimension, as mentioned earlier, the classification can be strategic, operational or tactical. Another possibility is the PWC (2004) classification as defined below, which is mainly for the financial domain.

1. Operational: Spreadsheets used to facilitate tracking and monitoring of workflow to support operational processes.
2. Analytical/Management Information: Spreadsheets used to support analytical review and management decision making. These may be used to evaluate the reasonableness of financial amounts.
3. Financial: Spreadsheets used to directly determine financial statement transaction amounts or balances that are populated into the general ledger and/or financial statements.

43

## 4. COMPARISON OF THE THREE MODELS:

The categorisation of the first model derived from the pilot study can be considered to be specific to this case study department. Considering the spectrum of use of spreadsheets within large organisations Model 1 seems the most appropriate.

The authors suggest that it might be complex to segregate into one of the 27 categories, therefore considering the practical application Model 2 was proposed. This model is simple to apply practically but it is weaker as there are two further possible dimensions (Complexity and Risk). This might need to be considered in organisations larger than the pilot organisation. Further this model is mainly for designing/development of new spreadsheets, and it does not take into account the thousands of existing spreadsheets which might be in use within big organisations. Even as far as the case study organisation is concerned, most of the spreadsheets tend to be in Section 1 of Model 2 (Figure 3). Within this department the level of urgency is low, because it concerns maintaining regular records and generating feedback at the end of year.

Model 3 uses a more understandable terminology within the majority of organisations. This aspect makes it easier to understand and implement in different types of organisations. Furthermore the authors believe that for the dimensions in this model it is easier to define criteria and define categories. Means of measuring is yet to be investigated.

## 5. CONCLUSION:

The spectrum of spreadsheet use within the pilot study organisation was very diverse. The research suggests that the first model proposed by the authors was simple but the key problem encountered within this model was the difficulty to classify a specific spreadsheet within one of the 27 categories of spreadsheets. But the supportive argument is that it does cover use of spreadsheets within organisation well.

The Model 2 on the other hand is also simple but easier to understand. Findings suggest that this model is also easy to apply generally, but one might need to consider two further dimensions (Risk and Complexity) when considering bigger organisations. Emphasis is placed on new spreadsheet development, and there is some fuzziness in dealing with thousands of spreadsheets already existing within the organisation.

According to the analysis the Model 3 seems to be the most acceptable for categorising spreadsheets. The primary reason is that it uses more conventional and universally understandable terminology and could be applied to a wide range of organisations. More specifically the dimensions are easier to understand, therefore the model is easier to implement. The authors perceive that it would be easier categorise using the Model 3 dimensions, but the means of measurement and categorisation are yet to be investigated. The next stage of the research is to develop clear definitions of each of the dimensions i.e. Magnitude and Dependence and then investigate appropriate criteria to measure and develop specific categories.

Blank Page